\documentclass[twocolumn,showpacs,preprintnumbers,amsmath,amssymb,prb]{revtex4}

 \def\XXint#1#2#3{{\setbox0=\hbox{$#1{#2#3}{\int}$}
     \vcenter{\hbox{$#2#3$}}\kern-.5\wd0}}

\def\fakebold#1{\relax\ifvmode\leavevmode\fi%
\ifmmode%
\setbox0=\hbox{$#1$}%
\else%
\setbox0=\hbox{#1}%
\fi%
\kern-.02em\copy0 \kern-\wd0%
\kern .04em\copy0 \kern-\wd0%
\kern-.0125em\raise.02em\box0%
}%

\usepackage{graphicx} \usepackage{dcolumn} \usepackage{bm}

\begin{document}



\title{Dispersion Instability in Strongly Interacting Electron
  Liquids}

\author{Ying Zhang} \author{Victor M. Yakovenko} \author{S. Das Sarma}
\affiliation{Condensed Matter Theory Center, Department of Physics,
  University of Maryland, College Park, MD 20742-4111}

\date{\today} 

\begin{abstract}
We show that the low-density strongly interacting electron liquid,
interacting via the long-range Coulomb interaction, could develop a
dispersion instability at a critical density associated with the
approximate flattening of the quasiparticle energy dispersion. At the
critical density the quasiparticle effective mass diverges at the
Fermi surface, but the signature of this Fermi surface instability
manifests itself away from the Fermi momentum at higher densities.
For densities below the critical density the system is unstable since
the quasiparticle velocity becomes negative. We show that one physical
mechanism underlying the dispersion instability is the emission of
soft plasmons by the quasiparticles. The dispersion instability occurs
both in two and three dimensional electron liquids.  We discuss the
implications of the dispersion instability for experiments at low
electron densities.
\end{abstract}

\pacs{71.10.-w; 71.10.Ca; 73.20.Mf; 73.40.-c}

\maketitle

\section{Introduction}
\label{sec:intro}

Recent theoretical work~\cite{diverge} on quasiparticle properties of
interacting electron systems has indicated the possibility of a
divergence of the zero-temperature quasiparticle effective mass at a
critical density, both in two-dimensional (2D) and three-dimensional
(3D) quantum Coulomb systems. In this paper, we investigate the nature
of this mass divergence by considering the quasiparticle energy
dispersion over a finite wavevector range around $k_F$, obtaining the
interesting result that the effective mass divergence is not just a
property of the Fermi surface (i.e. it does not happen only at $k_F$),
but is a dispersion instability where the whole quasiparticle energy
dispersion around the Fermi surface becomes essentially almost flat --
in fact, the instability first occurs at wavevectors far away from
$k_F$ at densities well above the critical density with the mass
divergence eventually moving to the Fermi surface at the critical
density. For densities lower than the critical density for mass
divergence, we find that interaction effects drive the quasiparticle
energy dispersion `concave' instead of the usual convex parabolic
energy dispersion of noninteracting free electrons, i.e., the energy
actually decreases with increasing wavevector implying a `negative'
effective mass -- the electrons have a `negative velocity' around the
Fermi surface. Thus, the effective mass divergence within the ring
diagram approximation for the electron self-energy reported in
Ref.~\onlinecite{diverge} is actually a more general dispersion
instability (i.e. the tendency of the quasiparticle dispersion of
developing a flat band around $k_F$) of the type first discussed by
Khodel and Shaginyan~\cite{khodel1} -- the `fermionic condensate'
formation in the terminology of Ref.~\onlinecite{khodel1}, fifteen
years ago.

We consider the standard jellium model for an electron system with the
electron-electron interaction being the usual `$1/r$' long-range
Coulomb interaction (in 2D or 3D) and the noninteracting kinetic
energy being the usual parabolic dispersion with a bare mass `$m$'.
Such a system is characterized~\cite{book} by a dimensionless
interaction parameter, the so-called Wigner-Seitz radius $r_s \equiv
(\pi n)^{-1/2}/a_B$ (2D); $(4 \pi n /3)^{-1/3} /a_B$ (3D), where $n$
is the 2D or 3D electron density and $a_B = \hbar^2 /(m e^2)$ is the
Bohr radius -- $r_s$ is both the effective interparticle separation
measured in the units of Bohr radius and the ratio of the average
Coulomb potential energy to the noninteracting kinetic energy. It was
shown in Ref.~\onlinecite{diverge} that the on-shell quasiparticle
effective mass, $m^*$, diverges (both in 2D and 3D) when the
quasiparticle mass renormalization is calculated within the
leading-order approximation in the dynamically screened Coulomb
interaction, or equivalently, in the infinite series of ring diagrams
for the reducible polarizability function. The critical $r_s$-value
($r_s^*$) for this effective mass divergence was found to be $r_s^*
\approx 16$ (2D); $48$ (3D). It was argued in
Ref.~\onlinecite{diverge} that, although the specific value $r_s^*$ is
surely model-dependent (and one cannot expect the ring diagram
approximation to give an exact or perhaps even an accurate value for
the critical $r_s$), the qualitative fact that there is a
quasiparticle effective mass divergence in strongly interacting 2D and
3D quantum Coulomb plasmas is a generic feature independent of the
details of the approximation. In Ref.~\onlinecite{diverge} it was
speculated that this effective mass divergence is the continuum analog
of the Mott transition, or equivalently, a precursor to the Wigner
crystallization (or perhaps a charge density wave instability).

The other possibility is that this effective mass
divergence~\cite{diverge} is the ``fermionic condensation'' discussed
in Ref.~\onlinecite{khodel1}, and further elaborated, developed and
discussed in Refs.~\onlinecite{khodel2, nozieres, khveshchenko,
  volovik, zverev, lidsky}. The ``fermionic condensate'' idea,
pioneered by Khodel and collaborators, involves a flattening of the
quasiparticle dispersion around $k_F$ at a critical value of the
interaction parameter, leading to interesting possibilities for the
Fermi distribution function and the Fermi liquid theory. The inherent
energy degeneracy associated with such a `band flattening', where
quasiparticles at different wavevectors have the same quasiparticle
energy, could lead to various instabilities, and the `renormalized
degenerate' interacting system could in principle reorganize itself
into a new (and perhaps an `exotic') phase such as a superconducting
condensate~\cite{khodel2, nozieres} or a charge density
wave~\cite{khodel3}. In such a situation involving a quantum phase
transition to a non-Fermi liquid phase, the effective mass divergence
is a signature of the emergence of the new collective phase, and the
energy degeneracy associated with the band flattening (or
equivalently, the plateau formation in the energy-momentum dispersion)
within the Fermi liquid phase itself is not a particularly critical
issue (except as a precursor to the eventual quantum phase
transition). The `negative velocity' (i.e. decreasing energy with
increasing momentum) that we find in our theory for $r_s > r_s^*$
(i.e.  below the critical density) obviously indicates an instability
of the Fermi liquid, and our goal in this paper is to better
understand this phenomenon. Note that the electron-electron
interaction we use in our model of 2D or 3D electron liquids is the
realistic long-range Coulomb interaction.

Since the effective mass divergence theoretically discovered in
Ref.~\onlinecite{diverge} involves a very specific and extremely
well-understood many-body approximation~\cite{book, rice}, namely the
ring-diagram approximation (sometimes referred to as the RPA
self-energy calculation or equivalently as the GW
approximation~\cite{hedin}) or its simple
generalizations~\cite{hubbard} incorporating approximate vertex
corrections through local field corrections, we focus in the current
work on a detailed calculation (within the same ring diagram
approximation) of the full quasiparticle energy dispersion $E({\bf
  k})$ to understand the nature of the mass divergence by
investigating whether the generalized wavevector-dependent
quasiparticle effective mass, $m({\bf k}) \equiv [ (\hbar k)^{-1} d
E({\bf k}) / d k]^{-1}$, diverges only at $k = k_F$ or the
quasiparticle dispersion is affected more drastically for a finite
range of $k$ around $k_F$ akin to the band-flattening phenomenon
leading to quasiparticle energy degeneracy. We find the latter
situation to be the case with the whole energy dispersion being
drastically affected by interaction effects at large $r_s$. We also
find that within our leading-order self-energy calculation (in the
dynamically screened Coulomb interaction) the dispersion instability
and the associated effective mass divergence arise, at least
partially, from the emission of collective modes (``plasmons'') by the
quasiparticles, which could happen at rather low wavevectors ($k
\gtrsim k_F$) in the strongly interacting regime of large $r_s$ (where
the mass divergence phenomenon occurs). Our identification of a
specific physical mechanism contributing partially to the mass
divergence and the dispersion instability phenomena, namely the
spontaneous emission of collective plasmon excitations by the
quasiparticles leading to a negative quasiparticle velocity (or
equivalently, a strong nonlinear ``dip'' in the quasiparticle
dispersion), suggests that the dispersion instability and the
effective mass divergence is a generic feature of strongly interacting
quantum Coulomb electron systems and is not a artifact of the ring
diagram approximation.

The rest of this paper is organized as follows. In \ref{sec:theory} we
describe the theory and formalism; in \ref{sec:results} we present our
results for the quasiparticle energy dispersion; and we conclude in
\ref{sec:con} giving a detailed discussion of the implications of our
results.

\section{Theory and formalism}
\label{sec:theory}

In this section we present the formalism we are going to use in our
work. Our goal is to calculate the quasiparticle energy dispersion as
a function of the interaction parameter $r_s$. The central quantity
needed for this calculation is the electron self-energy function.
Without any loss of generality we assume the electron system to be
spinless for our calculations since spin plays no role in the theory
as Coulomb interaction is spin independent. We choose $\hbar = 1$
throughout, which makes wavevector and momentum (as well as energy and
frequency) equivalent.


\subsection{Random Phase Approximation}

\begin{figure} [htbp]
\centering \includegraphics[width=3in]{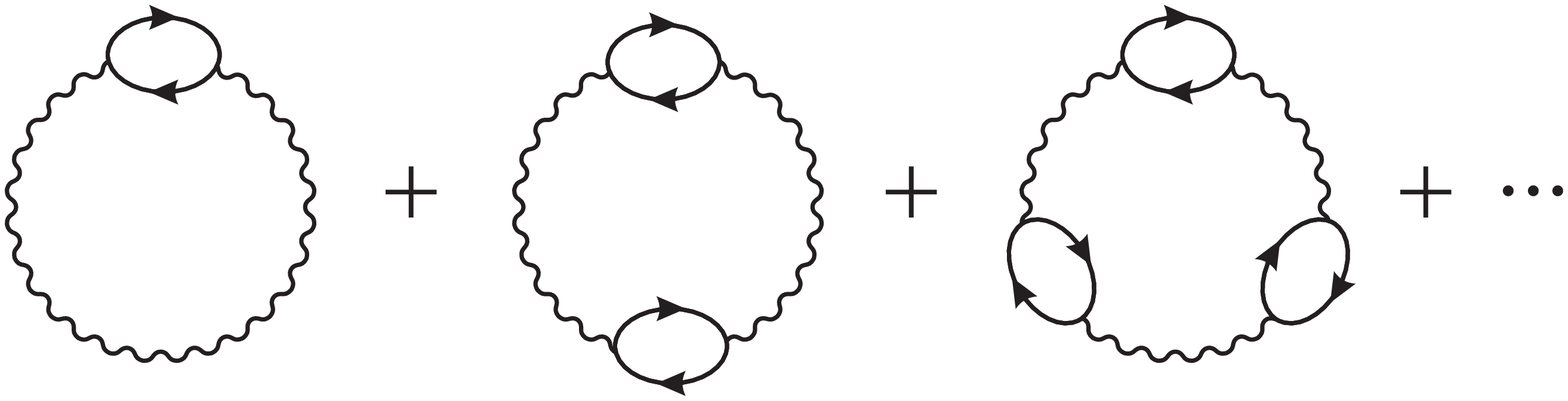}
\caption{The Feynman diagram for the Coulomb interaction contribution
  to the ground state energy within RPA. The circles are polarization
  bubbles, the wiggly lines are the bare Coulomb interaction, and the
  solid lines the noninteracting electron Green's function.}
  \label{fig:ground}
\end{figure}

We examine the jellium electron system with long-range Coulomb
interaction between electrons at zero temperature. Within random phase
approximation (RPA)~\cite{book, rice, hedin, hubbard}, the part of the
ground state energy introduced by Coulomb interaction can be denoted
by the Feynman diagrams shown in Fig.~\ref{fig:ground}. Following
Landau's approach, the quasiparticle energy can be obtained by
\begin{equation}
\label{eq:quasiE}
E_{\bf k} = {\delta E_G \over \delta n_{\bf k}},
\end{equation}
where $n_{\bf k}$ is the distribution function at momentum ${\bf k}$
and $E_G$ is the ground state energy of the system.

\begin{figure} [htbp]
\centering \includegraphics[width=3in]{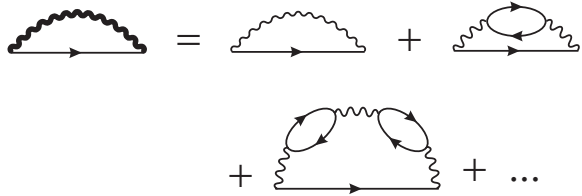}
\caption{The Feynman diagram for the self-energy within RPA. The thick
  wiggly line denotes the dynamically screened Coulomb interaction.}
  \label{fig:self}
\end{figure}

Note that the distribution function enters into the expression of
ground state energy $E_G$ through the noninteracting Green's function
\begin{equation}
\label{eq:G0}
G^{(0)} ({\bf k}, \omega) = 
{n_{\bf k} \over \omega - \xi_{\bf k} - i \eta} +
{1 - n_{\bf k} \over \omega - \xi_{\bf k} + i \eta},
\end{equation}
where $\xi_{\bf k} = k^2/(2m) - E_F$ is the non-interacting electron
energy with $E_F$ as the Fermi energy or the non-interacting chemical
potential, and $\eta$ is an infinitesimal positive number. It is easy
to obtain the variational derivative of the Green's function as
\begin{equation}
\label{eq:dG}
{\delta \over \delta n_{\bf k}} G^{(0)} ({\bf q}, \omega)
= 2 \pi i \delta (\omega - \xi_{\bf k}) \delta ({\bf k - q}).
\end{equation}
Graphically, taking the $n_{\bf k}$ variational derivative of a
quantity simply means cutting one solid line of the Feynman diagram
and taking the external momentum and frequency to be on-shell (i.e.
$\omega = \xi_{\bf k}$). We show the corresponding Feynman diagram for
the self-energy in Fig.~\ref{fig:self}. We emphasize that the RPA or
the ring-diagram approximation (which is appropriate for electron
liquids interacting with the long-range Coulomb interaction) as shown
in Fig.~\ref{fig:ground} necessarily implies that the on-shell
self-energy approximation is used for calculating the quasiparticle
energy dispersion (Fig.~\ref{fig:self}) since all energy and momenta
in Fig.~\ref{fig:ground} correspond to the noninteracting system.

\begin{figure} [htbp]
\centering \includegraphics[width=3in]{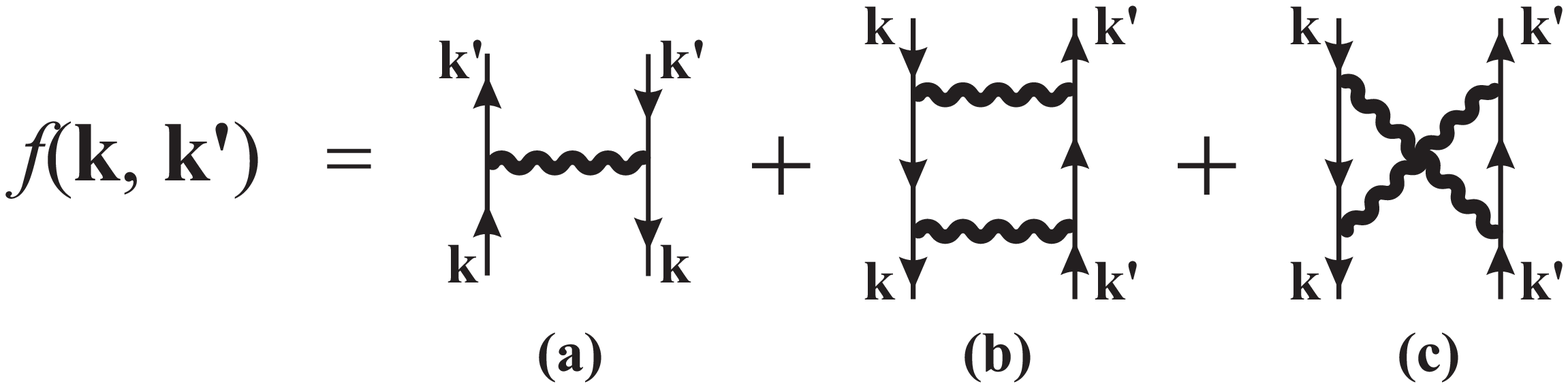}
\caption{The Feynman diagram for the Landau interaction function $f
  ({\bf k}, {\bf k'})$ within RPA.}
  \label{fig:landau}
\end{figure}

The second order derivative of the total ground energy is referred to
as Landau's interaction function:
\begin{equation}
\label{eq:f}
f({\bf k}, {\bf k'}) = {\partial^2 E_G \over 
\partial n_{\bf k} \partial n_{\bf k}}. 
\end{equation}
The Feynman diagram for the interaction function within RPA is shown
in Fig.~\ref{fig:landau}.

Landau's interaction function (\ref{eq:f}) determines renormalization
of the effective mass $m^*$ relative to the bare mass $m$
\begin{equation}
\label{eq:m-m*-f}
{1 \over m^*}={1 \over m} - C \int f(\theta) \cos\theta\,do,
\end{equation}
where $\theta$ is the angle between ${\bf k}$ and ${\bf k'}$, $do$ the
element of solid angle along ${\bf k'}$ in 3D and $d\theta$ in 2D, and
$C=k_F/(2\pi)^3$ in 3D and $C=1/(2\pi)^2$ in 2D.

The contribution (a) in Fig.\ \ref{fig:landau} to Landau's interaction
function is nothing but the static screened Coulomb interaction
\begin{equation}
\label{eq:static}
f_a(q)=-\frac{2\pi e^2}{q+\kappa} \quad \mbox{in 2D}, \quad
f_a(q)=-\frac{4\pi e^2}{q^2+\kappa^2} \quad \mbox{in 3D},
\end{equation}
where $q=|{\bf k}-{\bf k'}|$ is momentum transfer between the two
interacting electrons, and $\kappa$ is the appropriate inverse
screening length.  The minus sign in Eq.\ (\ref{eq:static}) reflects
the exchange character of the Coulomb interaction between two
fermions.  Because the amplitude of $f_a(q)$, given by Eq.\ 
(\ref{eq:static}), is maximal at $q=0$, i.e.\ at $\theta=0$, where
$\cos\theta>0$, it produces a positive contribution to the right-hand
side of Eq.\ (\ref{eq:m-m*-f}) and tends to decrease $m^*$.  Because
the first term $f_a$ dominates over the other two terms $f_b$ and
$f_c$ in Fig.\ \ref{fig:landau} at small $r_s$, thus $m^*<m$ at small
$r_s$ (see Fig.\ 6 of Ref.~\onlinecite{rice}).

However, the sum of the $f_b$ and $f_c$ terms in Fig.\ 
\ref{fig:landau} has the same sign as $f_a$ in Eq.\ (\ref{eq:static}),
but its maximal amplitude is achieved at ${\bf k}=-{\bf k'}$, i.e.\ at
$\theta=\pi$, where $\cos\theta<0$ (see Fig.\ 10 of
Ref.~\onlinecite{rice}).  Thus, $f_b$ and $f_c$ produce a negative
contribution to the right-hand side of Eq.\ (\ref{eq:m-m*-f}) and tend
to increase $m^*$.  Because the strength of the $f_b$ and $f_c$ terms
(``two-plasmon exchange'') increases relative to $f_a$ (``one-plasmon
exchange'') with the increase of $r_s$, the effective mass $m^*$
starts to increase and becomes greater than $m$.  This effect was
already seen in Fig.\ 6 of Ref.~\onlinecite{rice}.  However, the
numerical calculations of Ref.~\onlinecite{rice} stopped at a moderate
$r_s\sim 4$.  In Ref.~\onlinecite{diverge}, essentially the same
calculations were extended to greater $r_s$, and divergence of $m^*$
was found.  The divergence of $m^*$ occurs at the point where the
interaction term in the right-hand side of Eq.\ (\ref{eq:m-m*-f})
becomes equal to the bare term.

An alternative, but in some sense similar scenario of mass divergence
was proposed in Ref.~\onlinecite{khodel3} based on a phenomenological
assumption that $f(q)$ has maximal amplitude at $q\sim2k_F$, where
$\cos\theta<0$, due to proximity to a density-wave instability.


\subsection{Dielectric function}

The key quantity in evaluating the self-energy is the dynamically
screened Coulomb interaction $u({\bf q}, \omega) \equiv v_q /
\epsilon({\bf q}, \omega)$, where $v_q$ is the Coulomb interaction in
momentum space and $\epsilon({\bf q}, \omega)$ is the dynamical
dielectric function~\cite{book, rice, hedin, hubbard}. The Coulomb
interaction is given by $v_q = 2 \pi e^2/ q$ (2D) and $4 \pi e^2 /
q^2$ (3D) with $q \equiv |{\bf q}| $ being the appropriate 2D or 3D
momentum. The dielectric function is given by the infinite geometric
series of the noninteracting ring diagrams where each ring is just the
noninteracting electron polarizability. The 2D and 3D dielectric
functions $\epsilon(q, \omega) = 1 - v_q \Pi(q, \omega)$ are given in
Stern~\cite{stern} and Linhard~\cite{lindhard} respectively, where
$\Pi(q, \omega)$ is the polarizability, which can be denoted as one
ring or bubble as in Fig.~\ref{fig:ground} and Fig.~\ref{fig:self}.

In actual calculations it is conventional to express all the
expressions in terms of the dimensionless units $r_s$. The relation
between $r_s$ and Fermi vector is $ r_s = 1/(\alpha k_F a_B)$, where
$\alpha^{(2D)} = 1/2$ and $\alpha^{(3D)} = (9 \pi /2 )^{-1/3}$. For
completeness we write down the expression of 2D and 3D $\epsilon(q,
\omega)$ in the units of $2 k_F = 4 E_F = 2 m = 1$, i.e. we use $x =
q/(2k_F)$ to denote momentum and $u = \omega / (4 E_F)$ to denote
energy. For 2D $u > 0$ case we have:
\begin{eqnarray}
\label{eq:epsilon2D}
&&\!\!\!\!\!\!
\mbox{Re}~\epsilon(x, u) = 1 + {\alpha r_s \over 2 x}
\Bigg\{ 1 \nonumber \\
&& - {1 \over 2 x^2} \mbox{Sign} (x^2 - u) 
\Theta[(u - x^2)^2 - x^2] \sqrt{(u - x^2)^2 - x^2} \nonumber \\
&& 
- {1 \over 2 x^2} \mbox{Sign} (x^2 + u) 
\Theta[(u + x^2)^2 - x^2] \sqrt{(u + x^2)^2 - x^2} \Bigg\}, 
\nonumber 
\end{eqnarray}
\begin{eqnarray}
&&\!\!\!\!\!\!
\mbox{Im}~\epsilon(x, u) = {\alpha r_s \over 4 x^3} 
\Bigg\{ \Theta[x^2 - (u - x^2)^2] \sqrt{x^2 - (u - x^2)^2 } \nonumber
\\
&& - \Theta[x^2 - (u + x^2)^2 ] \sqrt{x^2 - (u + x^2)^2 } \Bigg\},
\nonumber
\end{eqnarray}
\begin{eqnarray}
&&\!\!\!\!\!\!
\epsilon(x, iu) = 1 + {\alpha r_s \over 2 x} 
\Bigg[1 \nonumber \\
&& -{1 \over \sqrt{2} x^2} \sqrt{x^4-x^2-u + \sqrt{(x^4 - x^2 - u)^2 +
    4 u^2  x^4} } \Bigg],
\end{eqnarray}
where $\Theta(x) = 1$ when $x > 0$ and $0$ otherwise.  For 3D $u > 0$
case we have
\begin{eqnarray}
\label{eq:epsilon3D}
\mbox{Re}~\epsilon(x, u) = 1 &+& {\alpha r_s \over \pi 4 x^2}
\Bigg\{ 1 \nonumber \\
&+& {1 \over 4 x^3} [x^2 - (x^2 + u)^2] 
\ln \left|{q^2+q+u \over q^2-q+u} \right| \nonumber \\
&+& {1 \over 4 x^3} [x^2 - (x^2 - u)^2] 
\ln \left|{q^2+q-u \over q^2-q-u} \right| \Bigg\}
\nonumber 
\end{eqnarray}
\begin{eqnarray}
\mbox{Im}~\epsilon(x, u) = \left\{\begin{array}{ll} 
{\alpha r_s u \over 2 x^3} & (x-x^2>u>0) \\ \\
{\alpha r_s \over 8 x^5} [x^2 - (x^2 - u)^2] 
& (x^2+x > u > |x^2 -x|) \\ \\
0 & (\mbox{Otherwise}) 
\end{array} \right. \nonumber 
\end{eqnarray}
\begin{eqnarray}
\epsilon(x, iu) = 1 &+& {\alpha r_s \over \pi 4 x^2}
\Bigg\{1 \nonumber \\
&+& {1 \over 4 x^3} [x^2 - x^4 + u^2]
\ln \left[{ (q^2 + q)^2 + u^2 \over (q^2 - q)^2 + u^2} \right]
\nonumber \\
&+& {u \over x} 
\left[\arctan ({u \over q^2+q}) - \arctan ({u \over q^2-q}) \right]
\Bigg\}.   
\end{eqnarray}
In both 2D and 3D we have the relation $\epsilon(x, -u) =
\epsilon^*(x, u)$ (complex conjugate) and $\epsilon(x, -iu) =
\epsilon(x, iu)$.


\subsection{Quasiparticle self-energy}

The one-loop self-energy (Fig.~\ref{fig:self}) at zero temperature can
be written as
\begin{eqnarray}
\label{eq:E}
\Sigma ({\bf k}, \omega) &=& - \int {d^d q \over (2 \pi)^d}
\int {d \nu \over 2 \pi i} {v_q \over \epsilon({\bf q}, \nu)}
\nonumber \\
&&~~~~~~~~ \cdot G^{(0)}({\bf k+q}, \omega + \nu).
\end{eqnarray}

\begin{figure}[htbp]
\centering \includegraphics[width=2in]{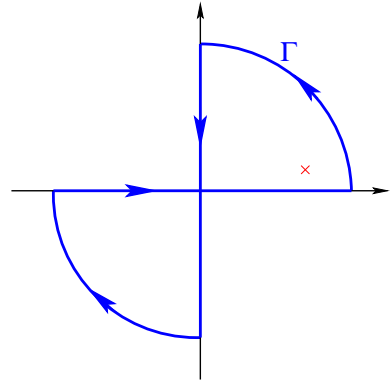}
  \caption{Contour in the frequency $\nu$ plane. The cross denotes the
    possible position of the pole of Green's function.}
\label{fig:contour}
\end{figure}

Due to the difficulty with the principal value integration and
singularities of $1 / \epsilon({\bf k}, \omega)$ along the real axis
in Eq.~(\ref{eq:E}), it is advantageous to follow the standard
procedure and deform the frequency integration from the real axis to
the imaginary axis. After choosing the contour of frequency
integration as in Fig.~\ref{fig:contour}, we consider the integration
\begin{equation}
\label{eq:contour}
\oint_{\Gamma} {d \nu \over 2 \pi i} w({\bf q}, \nu) 
G_0 ({\bf q} + {\bf k}, \nu + \omega),
\end{equation}
where $w({\bf q}, \nu)$ is any complex function that is analytic in
the upper and lower half of the complex plane which satisfies the
condition that $w({\bf q}, \nu) \to 0$ as $|\nu| \to \infty$. On the
one hand, (\ref{eq:contour}) equals the integration of the integrand
along real and imaginary axes since the integration along the curved
part of the contour $\Gamma$ (Fig.~\ref{fig:contour}) vanishes. On
the other hand, (\ref{eq:contour}) equals the residue due to the pole
of Green's function
\begin{eqnarray}
\label{eq:residue}
&-&[1 - n_F(\xi_{{\bf q} + {\bf k}}) ] 
\Theta(\omega - \xi_{{\bf q} + {\bf k}} )  
w({\bf q}, \xi_{{\bf q} + {\bf k}} - \omega - i \eta) \nonumber \\
&+& n_F(\xi_{{\bf q} + {\bf k}}) 
\Theta(\xi_{{\bf q} + {\bf k}} - \omega) 
w({\bf q}, \xi_{{\bf q} + {\bf k}} - \omega + i \eta),
\end{eqnarray}
where $n_F(x)$ is the Fermi distribution function (At $T=0$ $n_F(x) =
1$ for $x < 0$ and $0$ otherwise). Now by setting
\begin{equation}
\label{eq:w}
w({\bf k}, \omega) = {1 \over \epsilon({\bf k}, \omega)} -1,
\end{equation}
we have
\begin{eqnarray}
\label{eq:E1}
&& \!\!\!\!\!\!\!\!\!\!\!\!
\Sigma({\bf k}, \omega) = 
- \int {d^d q \over (2 \pi)^d} v_q n_F(\xi_{{\bf q} + {\bf k}}) 
\nonumber \\
&&- \int {d^d q \over (2 \pi)^d} 
\int {d \nu \over 2 \pi i} v_q w({\bf q}, \nu) 
G_0 ({\bf q} + {\bf k}, \nu + \omega) 
\nonumber \\
&=& 
- \int {d^d q \over (2 \pi)^d} v_q n_F(\xi_{{\bf q} + {\bf k}}) \\
&& - \int {d^d q \over (2 \pi)^d} 
\Big\{ - [ 1 - n_F(\xi_{{\bf q} + {\bf k}}) ] \nonumber \\
&&~~~ \cdot \Theta(\omega - \xi_{{\bf q} + {\bf k}} ) 
w({\bf q}, \xi_{{\bf q} + {\bf k}} - \omega - i \eta)  
\nonumber \\ 
&& ~~~~ + n_F(\xi_{{\bf q} + {\bf k}}) 
\Theta(\xi_{{\bf q} + {\bf k}} - \omega) 
w({\bf q}, \xi_{{\bf q} + {\bf k}} - \omega + i \eta) 
\Big\} \nonumber \\
&& - \int {d^d q \over (2 \pi)^d} \int {d \nu \over 2 \pi}
{w({\bf q}, i \nu) \over i \nu + \omega - \xi_{{\bf q} + {\bf k}}}.
\end{eqnarray}
Note that the first term in Eq.~(\ref{eq:E1}), usually named the
exchange part, is a singular term when temperature is zero. Thus we
rewrite the first and the second term so that they are well defined.
Real and imaginary parts of the self-energy $\Sigma({\bf k}, \omega) =
\Sigma'({\bf k}, \omega) + i \Sigma''({\bf k}, \omega)$ can then be
written as
\begin{eqnarray}
\label{eq:ReE}
&& \!\!\!\!\!\!\!\!\!\!
\Sigma'({\bf k}, \omega) = 
- \int {d^d q \over (2 \pi)^d} v_q 
\Theta(2 m \omega + k_F^2 - |{\bf q - k}|^2) \nonumber \\
&& + \int {d^d q \over (2 \pi)^d} 
v_q \mbox{Re}~{1 \over \epsilon({\bf q}, \xi_{\bf q - k} - \omega)} 
\nonumber \\
&&~~~\cdot \Big[\Theta(2 m \omega + k_F^2 - |{\bf q - k}|^2) 
- \Theta(k_F^2 - |{\bf q - k}|^2) \Big]\nonumber \\
&& - \int {d^d q \over (2 \pi)^d} \int {d \nu \over 2 \pi}
v_q \left[ {1 \over \epsilon({\bf q}, i \nu)} -1 \right]
{1 \over i \nu + \omega - \xi_{{\bf q} + {\bf k}}},
\end{eqnarray}

\begin{eqnarray}
\label{eq:ImE}
&&\Sigma''({\bf k}, \omega) = 
\int {d^d q \over (2 \pi)^d} 
v_q \mbox{Im}~{1 \over \epsilon({\bf q}, \xi_{\bf q - k} - \omega)} 
\nonumber \\
&&~~~\cdot \Big[\Theta(2 m \omega + k_F^2 - |{\bf q - k}|^2) 
- \Theta(k_F^2 - |{\bf q - k}|^2) \Big].
\end{eqnarray}

To be consistent with our leading-order one-loop approximation in the
dynamically screened Coulomb interaction (Figs.~\ref{fig:ground} and
\ref{fig:self}) we should calculate the self-energy only within the
on-shell approximation~\cite{diverge, rice} instead of solving the
full Dyson's equation. After putting $\omega = \xi_{\bf k}$ (i.e. the
on-shell approximation), we express the above equations in terms of
$r_s$, while using $2 k_F$ as the unit of wave-vector, and $4E_F$ as
the energy unit. For 2D real part of self-energy $\Sigma_{\bf k}
\equiv \Sigma({\bf k}, \omega = \xi_{\bf k})$ we obtain
\begin{eqnarray}
\label{eq:Er2D}
&& \!\!\!\!\!\!\!\!\!\!
{\Sigma'_{\bf k} \over 4 E_F }=
- {2 \alpha r_s \over \pi} y \nonumber \\
&& + {\alpha r_s \over \pi} \int_{1/2}^y dx \int_0^\pi d \theta 
{x \over \sqrt{x^2 - 2xy \cos(\theta) + y^2}}
\nonumber \\
&&~~~ \cdot
\mbox{Re}~{1 \over \epsilon 
\left( \sqrt{x^2 - 2xy \cos(\theta) + y^2}, x^2-y^2 \right) }
\nonumber \\
&& - {\alpha r_s \over \sqrt{2} \pi } \int d x
\int d u  \left[ {1 \over \epsilon({\bf x}, i x u)} -1 \right]
\nonumber \\
&&\cdot{ \sqrt{ (x^2 - u^2 - 4 y^2) 
+ \sqrt{ (x^2 - u^2 - 4 y^2)^2 + 4 y^2 u^2 } }
\over \sqrt{ (x^2 - u^2 - 4 y^2 )^2 + 4 x^2 u^2 } },
\end{eqnarray}
where ${\bf y} = {\bf k} / (2 k_F)$. Similarly for 3D we have
\begin{eqnarray}
\label{eq:Er3D}
&& \!\!\!\!\!\!\!\!\!\!
{\Sigma'_{\bf k} \over 4 E_F }=
- {\alpha r_s \over \pi} y \nonumber \\
&& + {\alpha r_s \over \pi} \int_{1/2}^y dx \int_0^\pi d \theta 
{x^2 \over x^2 - 2xy \cos(\theta) + y^2}
\nonumber \\
&&~~~ \cdot
\mbox{Re}~{1 \over \epsilon 
\left( \sqrt{x^2 - 2xy \cos(\theta) + y^2}, x^2-y^2 \right) }
\nonumber \\
&& - {\alpha r_s \over 4 \pi^2 y} \int_0^\infty d x
\int_0^\infty d u  \left[ {1 \over \epsilon({\bf x}, i x u)} -1 \right]
\nonumber \\
&&~~~\cdot
\ln \left[ {(2y - x)^2 + u^2 \over (2y + x)^2 + u^2} \right].
\end{eqnarray}
The 2D imaginary self-energy can be written down as,
\begin{eqnarray}
\label{eq:ImEr2D}
{\Sigma''_{\bf k} \over 4 E_F} &=&
{\alpha r_s \over \pi} \int_{1/2}^y dx \int_0^\pi d \theta 
{x \over \sqrt{x^2 - 2xy \cos(\theta) + y^2}} \nonumber \\
&& \cdot
\mbox{Im}~{1 \over \epsilon 
\left( \sqrt{x^2 - 2xy \cos(\theta) + y^2}, x^2-y^2 \right) },
\end{eqnarray}
and for 3D imaginary quasiparticle self-energy we have
\begin{eqnarray}
\label{eq:ImEr3D}
{\Sigma''_{\bf k} \over 4 E_F} &=&
{\alpha r_s \over \pi} \int_{1/2}^y dx \int_0^\pi d \theta 
{x^2 \over x^2 - 2xy \cos(\theta) + y^2}
\nonumber \\
&& \cdot
\mbox{Im}~{1 \over \epsilon 
\left( \sqrt{x^2 - 2xy \cos(\theta) + y^2}, x^2-y^2 \right) }.
\end{eqnarray}


\subsection{Quasiparticle energy dispersion and damping}

The quasiparticle dispersion is given simply by adding the on-shell
real self-energy to the noninteracting electron energy
\begin{equation}
\label{eq:EkDefine}
E_{\bf k} \equiv \xi_{\bf k} + \Sigma'_{\bf k}.
\end{equation}
The quasiparticle damping rate, which is non-zero away from the Fermi
surface $k \ne k_F$, is given by the on-shell imaginary part of the
self-energy
\begin{equation}
\label{eq:GammakDefine}
\Gamma_{\bf k} 
= \left| \mbox{Im}~\Sigma({\bf k}, \xi_{\bf k}) \right|
= \Sigma''_{\bf k}.
\end{equation}
The renormalized quasiparticle effective mass is given by
\begin{equation}
\label{eq:massDefine}
{1 \over m^* ({\bf k}) } = {1 \over k} {d E_{\bf k} \over d k}.
\end{equation}
It is clear that we get $m^* \equiv m$, the constant bare electron
mass, if we use the noninteracting energy $\xi_{\bf k} = k^2 / (2m)$
for $E_{\bf k}$, i.e. if we put the self-energy to be zero.  Putting
$k = k_F$ one gets the quasiparticle effective mass at the Fermi
surface, as calculated in Ref.~\onlinecite{diverge}. For an arbitrary
$k \ne k_F$, $m({\bf k})$ defines the dispersing quasiparticle mass,
i.e.  the generalized wavevector-dependent quasiparticle effective
mass.


\section{Results}
\label{sec:results}

In this section we present our calculated results showing the
quasiparticle energy dispersion for both 2D and 3D electron systems
with long range Coulomb interactions. The dispersion instability at
large $r_s$ will be obvious in these results.

\begin{figure}[htbp]
\centering \includegraphics[width=3.5in]{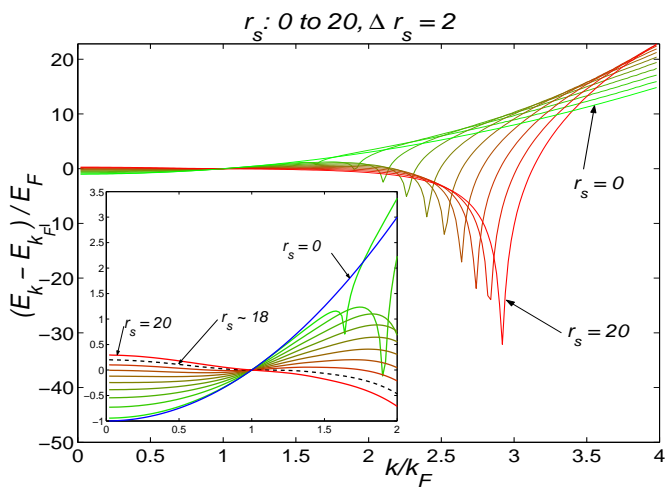}
  \caption{(Color online) 2D quasiparticle energy as a function of 
    momentum at $r_s$ values ranging from $0$ to $20$. Inset: zoom-in
    of the spectrum with $k/k_F < 2$. The dashed line indicate the
    $r_s$ value at which the effective mass at the Fermi surface shows
    divergence ($r_s^* \sim 18$).}
\label{fig:Ek2D1}
\end{figure}

\begin{figure}[htbp]
\centering \includegraphics[width=3.5in]{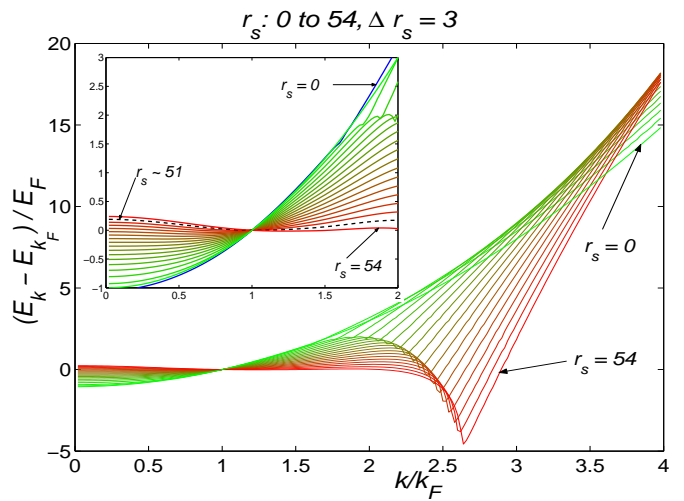}
  \caption{(Color online) 3D quasiparticle energy as a function of 
    momentum at $r_s$ values ranging from $0$ to $54$. Inset: zoom-in
    of the spectrum with $k/k_F < 2$. The dashed line indicate the
    $r_s$ value at which the effective mass at the Fermi surface shows
    divergence $r_s^* \sim 50$.}
\label{fig:Ek3D1}
\end{figure}

In Fig.~\ref{fig:Ek2D1} we plot the calculated quasiparticle energy as
a function of momentum for different $r_s$ values in a 2D system. It
is clear that the quasiparticle effective mass ($m^* = k_F (d E_k /d
k)^{-1}|_{k = k_F}$) diverges at some low density at critical $r_s$
value $r_s^* \sim 18$. As $r_s$ increases, not only the energy
dispersion of the quasiparticles around Fermi surface becomes flatter
(i.e.  relatively independent of momentum), but the dispersion in the
whole momentum space is fundamentally changed.  The quasiparticle
energy dispersion in a 3D electron system is presented in
Fig.~\ref{fig:Ek3D1}, showing the dispersion ``flattening'' for a
range of wavevector $k$ at a critical $r_s^* \sim 50$.

\begin{figure}[htbp]
\centering \includegraphics[width=3.5in]{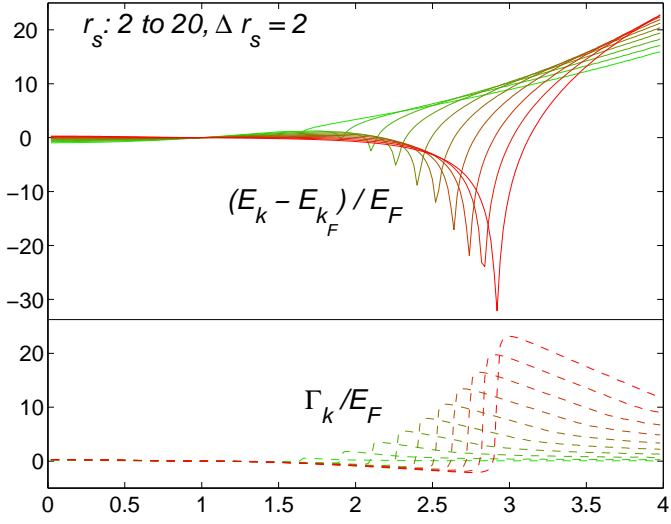}
  \caption{(Color online) 2D real and imaginary quasiparticle energy 
    as a function of momentum at $r_s$ values ranging from $2$ to
    $20$.}
\label{fig:ImE2D1}
\end{figure}

\begin{figure}[htbp]
\centering \includegraphics[width=3.5in]{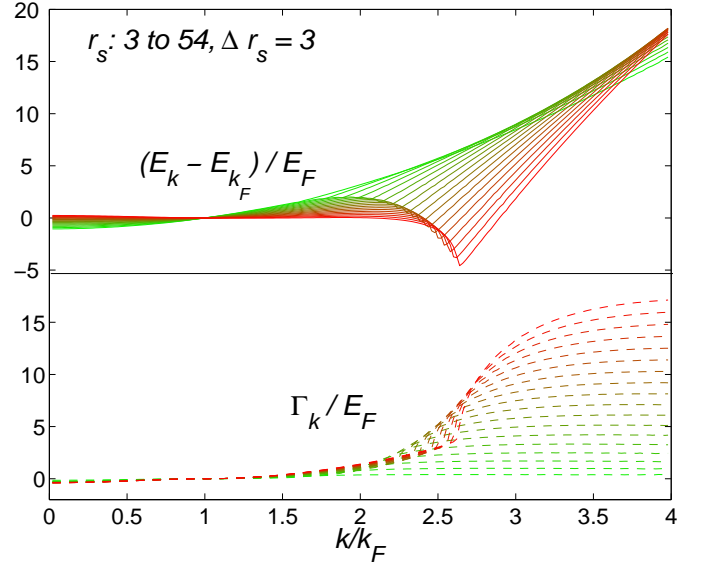}
  \caption{(Color online) 3D real and imaginary quasiparticle energy 
    as a function of momentum at $r_s$ values ranging from $3$ to
    $54$.}
\label{fig:ImE3D1}
\end{figure}

An obvious common feature of both 2D and 3D quasiparticle energy
dispersion is that at some large momentum, the dispersion shows a
kink, which indicates the plasmon-resonance. This is the threshold
wavevector at which the quasiparticle can emit plasmons obeying
energy-momentum conservation~\cite{jalabert}. This
plasmon-emission-induced kink in $E_{\bf k}$ for $k > k_F$ occurs at
all $r_s$, with the threshold wavevector $k_{th}$ being
$r_s$-dependent.  As $r_s$ increases, this resonance feature becomes
stronger. To further substantiate the connection between the
dispersion (and the associated effective mass divergence) and the
plasmon resonance we show in Figs.~\ref{fig:ImE2D1} (2D) and
\ref{fig:ImE3D1} (3D) our calculated quasiparticle energy dispersion
{\em together} with the calculated quasiparticle damping $\Gamma_{\bf
  k} \equiv Im \Sigma({\bf k}, \xi_{\bf k})$. It is obvious from
Figs.~\ref{fig:ImE2D1} and \ref{fig:ImE3D1} that the ``dip'' (at $k >
k_F$) in the quasiparticle dispersion occurs precisely at the plasmon
emission threshold wavevector~\cite{jalabert}. The energy-momentum
conservation makes it impossible~\cite{jalabert} for plasmon emission
by quasiparticles to happen below the threshold momentum
$k_{th}$. Since the real and imaginary parts of the electron
self-energy function are connected through the Kramers-Kronig
causality relations, the plasmon emission threshold in the
quasiparticle damping (i.e. $\mbox{Im} \Sigma$) shows up as a kink in
the quasiparticle dispersion (i.e.  $\mbox{Re} \Sigma$). This kink or
the dip in the dispersion indicates a nonlinear instability of the
quasiparticles over a finite range of momentum since the quasiparticle
velocity becomes manifestly negative for $k \lesssim k_{th}$
where the dispersion is locally inverted. With increasing $r_s$,
interaction effects become stronger and the plasmon-emission-induced
kink also becomes deeper indicating a progressively stronger (with
increasing $r_s$) nonlinear instability around $k_{th}$.  The
plasmon emission threshold $k_{th} (> k_F)$ is given by the
solution of the following transcendental equation:
\begin{equation}
\label{eq:kth}
\max_{k_F < q < k_{th}} 
\epsilon(k_{th} - q, {k^2_{th} \over 2m} 
- {q^2 \over 2m}) = 0.
\end{equation}

Below we discuss the (partial) connection between the plasmon emission
phenomenon (leading to the `kink' in the quasiparticle dispersion at
the plasmon resonance momentum $k_{th}$) and the dispersion
instability, which is quite apparent in Figs.~\ref{fig:Ek2D1},
\ref{fig:Ek3D1}, \ref{fig:ImE2D1}, \ref{fig:ImE3D1}, and at the same
time emphasize some subtle features about this connection, which point
to the qualitative nature of this connection.

\begin{figure}[htbp]
\centering \includegraphics[width=3.5in]{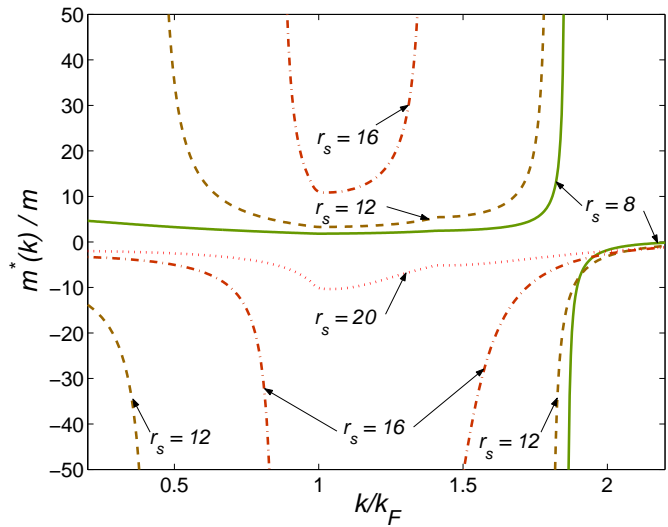}
  \caption{(Color online) Momentum dependent 2D quasiparticle
    effective mass for $r_s = 8, 12, 16, 20$.}
\label{fig:mk2D1}
\end{figure}

\begin{figure}[htbp]
\centering \includegraphics[width=3.5in]{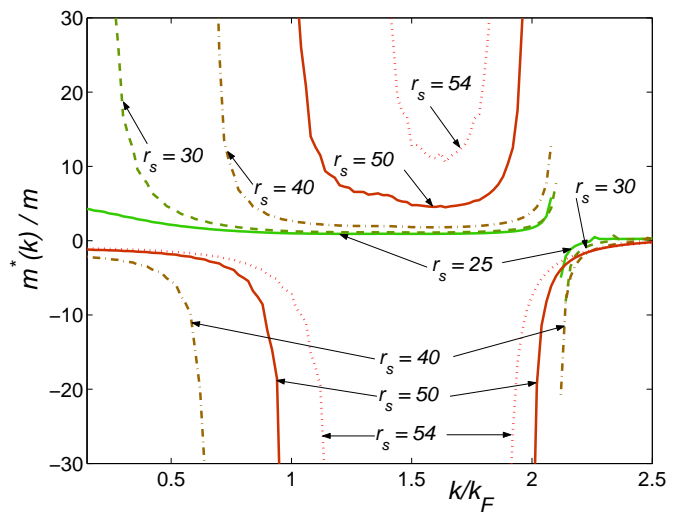}
  \caption{(Color online) Momentum dependent 3D quasiparticle
    effective mass for $r_s = 25, 30, 40, 50, 54$.}
\label{fig:mk3D1}
\end{figure}

A closer examination of the energy dispersion (see
Figs.~\ref{fig:Ek2D1} and \ref{fig:Ek3D1}) provides interesting
insights. Here we focus on the momentum region where the dispersion
shows instability. The divergence of the effective mass (i.e. the
inverse slope of the energy spectrum) for quasiparticles at different
wavevecter (not just at the Fermi surface) is a good indication of
these instabilities (also see Figs.~\ref{fig:mk2D1} and
\ref{fig:mk3D1}). Due to the presence of the plasmon resonance which
we described in the last paragraph, one effective mass divergence
happens at momentum (which we call the upper instability momentum
$k_h^*$) slightly smaller than the threshold resonance momentum
($k_{th}$) for both 2D and 3D systems, i.e. $m^*(k) \to \infty$
as $k \to k_h^*$. The energy dispersion reaches a local maximum when
$k = k_h^*$. As we mentioned, this feature exists for all $r_s$
values. However, for low electron density, this is not the only
instability feature that occurs. In fact in both 2D and 3D, for $r_s$
above a certain value, a {\em quasihole} instability emerges at a
smaller momentum, which we call the lower instability momentum
$k_l^*$. The energy dispersion reaches local minimum when $k = k_l^*$,
and $m^*(k) \to \infty$ as $k \to k_l^*$. In other words, the slope of
the energy dispersion is negative in the region $k < k_l^*$, positive
for $k_l^* < k < k_h^*$, negative for $k_h^* < k < k_{th}$, and
then positive for $k > k_{th}$. So it is clear that, for big
enough $r_s$ (but still smaller than $r_s^*$), there are two
instability regions $k < k_l^*$ and $k_h^* < k < k_{th}$ away
from Fermi surface ($k_l^* < k_F < k_h^*$), where the momentum
dependent effective mass is negative. (For very small $r_s$ we only
have the quasiparticle instability region $k_h^* < k <
k_{th}$.) We note that while the quasiparticle dispersion
instability at $k^*_h \lesssim k_{th}$ is induced by plasmon
emission due to the obvious connection between $k_h^*$ and
$k_{th}$, no such simple explanation seems to apply to the
corresponding quasihole instability at $k < k_l^*$.

For $r_s < r_s^*$, the lower and upper instability regions remain
below and above Fermi surface, and therefore are not of any particular
significance since the quasiparticle damping ($\Gamma_{\bf k} \equiv
\mbox{Im} \Sigma({\bf k}, \xi_{\bf k})$) is finite. As $r_s$
increases, both of these two regions grow larger: $k_l^*$ increases,
$k_h^*$ decreases, and $k_{th}$ increases with increasing
$r_s$. If one (or both) of these two instability regions reaches the
Fermi surface, the effective mass on the Fermi surface diverges.  This
is indeed what happens except that the details are slightly different
depending on the system dimensionality (2D or 3D). In 2D, as $r_s \to
r_s^*$, both instability regions reach $k_F$ at the same time, whereas
in 3D it appears that the quasihole instability reaches $k_F$ first at
$r_s = r_s^*$. The difference between 2D and 3D results arises from
the very different density of states in the two cases.

For 2D, when $r_s \sim 10$, $k_l^*$ starts out at zero momentum, and
as $r_s$ increases, $k_l^*$ increases and at the same time $k_h^*$
decreases.  As $r_s$ reaches the critical value $r_s^* \sim 18$,
$k_l^* = k_h^* = k_F$, which is the critical value at which the
effective mass at the Fermi surface diverges. When $r_s > r_s^*$,
there is no effective mass divergence in the whole momentum space, and
the dispersion is inverted for all $k < k_{th}$. The 3D case is
a little different. As $r_s \sim 28$, $k_l^*$ first emerges at zero
momentum, and keeps increasing as $r_s$ increases while $k_h^*$
decreases. As $r_s$ reaches the value $r_s^* \sim 50$, $k_l^* = k_F$,
and the effective mass at the Fermi surface diverges. However at this
$r_s$, another effective mass divergence happens at $k_h^* > k_F$.
Only when $r_s$ increases to more than $58$, $k_l^*$ and $k_h^*$ meet
each other at around $1.6 k_F$, after which the effective mass does
not diverge in the whole momentum space.

Because it is difficult to visually locate the local maximum and
minimum in the energy dispersion, we show in Figs.~\ref{fig:mk2D1} and
~\ref{fig:mk3D1} our calculated momentum-dependent quasiparticle
effective mass $m^*({\bf k}) = [k^{-1} d E_{\bf k} /d k]^{-1}$.  Of
course, the concept of a quasiparticle effective mass far away from
the Fermi surface is not particularly meaningful since these
quasiparticles are necessarily highly damped with short lifetimes.
Nevertheless it is important to realize that the effective mass
divergence initially develops for $r_s < r_s^*$ (i.e.  at densities
much larger than the critical density for the effective mass
divergence at the Fermi surface) at wavevectors $k_h^* > k_F$ and
$k_l^* < k_F$ far away from the Fermi surface. Eventually at $r_s =
r_s^*$ the effective mass divergence reaches $k_F$, becoming at the
same time a dispersion instability where the quasiparticle dispersion
$E_{\bf k}$ essentially becomes almost flat all around $k_F$, implying
$m({\bf k})$ to be divergent at the Fermi surface. This approximate
``band-flattening'', i.e. $E_{\bf k}$ approximately independent of
${\bf k}$, for large enough $r_s$ values is also apparent in
Figs.~\ref{fig:Ek2D1}, \ref{fig:Ek3D1}, \ref{fig:ImE2D1},
\ref{fig:ImE3D1}.

Before concluding this section we mention that the critical $r_s^*$
for the effective mass divergence ($18$ for 2D and $50$ for 3D)
obtained in this work is slightly larger than that obtained in
Ref.~\onlinecite{diverge} ($16$ in 2D and $48$ in 3D) since the
current work considers spinless (or equivalently, spin-polarized)
electrons whereas Ref.~\onlinecite{diverge} dealt with a paramagnetic
electron system with a spin degeneracy of $2$. It is interesting that
$r_s^*$ shows only a weak dependence on the spin-polarization
properties of the electron liquid. We believe that, while the basic
mass divergence and band flattening phenomena are generic phenomena in
strongly interacting electron liquids, the precise value of $r_s^*$
must depend on the approximation scheme involved, and it is likely
that the ``correct'' $r_s^*$ is larger than that obtained within RPA.
Our reason for using spinless electrons in our calculation is that
within RPA the electron system undergoes a ferromagnetic
transition~\cite{sus} of either Stoner or Bloch type to full spin
polarization at a critical $r_s$ value which is lower than $r_s^*$,
the critical $r_s$ for the mass divergence, and therefore it is more
appropriate to carry out the dispersion-instability/mass-divergence
calculation for a fully spin-polarized electron system as we do in the
current work. We also notice that previous theoretical
works~\cite{rajagopal, ortiz} predicted possible partial
spin-polarized ground state in 3D electron systems. We do not consider
this effect in our calculation because within the same approximation
scheme as in the present work, we find~\cite{sus} that in 3D electron
systems, the densities of the partial spin-polarization region are
much higher than the density that corresponds to the divergence of
effective mass, and therefore partial spin-polarization has no effect
on the dispersion instability. Partial spin-polarization does not
occur in 2D systems~\cite{rajagopal}, and thus this issue does not
arise for our 2D calculations.


\section{Discussion and conclusion}
\label{sec:con}

We show in this paper that the standard (and widely used) ring diagram
approximation (RPA) self-energy calculation leads to a quasiparticle
dispersion instability in the strongly interacting electron liquid
(both 2D and 3D) at a low critical density -- the quasiparticle
dispersion becomes almost ``flat'' around $k_F$ for $r_s = r_s^*$ and
beyond the critical point, for $r_s > r_s^*$, the quasiparticle
effective velocity becomes negative implying an instability. The band
flattening phenomenon is similar to that envisioned by Khodel and
Shaginyan~\cite{khodel1} in a different context some years ago. We
have also identified a physical mechanism leading to the dispersion
instability and the associated effective mass divergence.  We find
that it arises partially from plasmon (or collective mode) emission by
the quasiparticles -- in some sense, the ``recoil'' associated with
strong plasmon emission slows down the quasiparticles, eventually
effectively ``stopping'' it, leading to the mass divergence and band
flattening.  We emphasize however, that the plasmon emission and
recoil mechanism is at best a partial physical mechanism for the
dispersion instability -- it plays a crucial role, but is not the
whole story.

Two important questions immediately arise in the context of our
theoretical findings: (1) What does the dispersion instability signify
or imply? (2) What, if any, are its experimental implications and
consequences? We cannot answer either of these questions definitively.
But we can speculate some possibilities.  Since the answer to the
second question obviously depends on the answer to the first question,
we first discuss some possible answers to the first question.

The meaning or implication of the dispersion instability associated
with the effective mass divergence and the quasiparticle band
flattening has earlier been discussed in the literature. Two of us
have earlier speculated~\cite{diverge} that the effective mass
divergence is essentially a continuum version of the Mott transition
as envisioned, for example, in the Brinkman-Rice
scenario~\cite{brinkman}. Such a Brinkman-Rice Mott transition
scenario had earlier been invoked~\cite{vollhardt} in the context of
the ``almost localization'' phenomenon in normal He-3 where the
short-ranged inter-fermion interaction (in contrast to the long-ranged
Coulomb interaction we are considering) is known to lead to
quasiparticle mass divergence in the strongly interacting regime
(which curiously happens at very high, rather than very low, densities
in He-3 since the interaction is short-ranged in that system).
Usually the continuum analog of the Mott transition in an electron
liquid system is thought to be the Wigner crystallization transition,
which, according to the best current quantum Monte Carlo (QMC)
simulations, occur at a critical $r_s$ of $38$ (2D) and $90$ (3D). It
is entirely possible that the effective mass divergence and the band
flattening we find within RPA is the precursor to (or perhaps the RPA
signature for) the Wigner transition, but we cannot prove or establish
this speculation with any kind of theoretical arguments. It would be
highly desirable in this context to carry out QMC calculations to
study the dispersion instability, but QMC studies are reliable only
for the true ground state properties, and may be quite inaccurate for
effective mass calculations.

A second possibility, closely related to the Wigner crystallization
transition discussed above, is that the dispersion instability is
essentially a charge density wave instability~\cite{khodel3}. This is
not an absurd proposition given that the possibility of a
density-driven charge density wave instability in a jellium electron
liquid was originally discussed~\cite{overhauser} by Overhauser more
than forty years ago. We have therefore carefully analyzed the
wavevector dependent contributions to the quasiparticle self-energy as
well as calculated the Fermi liquid interaction function $f({\bf k},
{\bf k'})$ to see if there are any characteristic wavevectors which
predominantly contribute to the dispersion instability. As should be
obvious from our results, there are no characteristic wavevectors in
the effective mass divergence phenomenon that we have discovered since
it arises from a dispersion instability in which the whole dispersion
is strongly affected and all wavevectors seem to be equivalent. The
plasmon emission threshold wavevector $k_{th}$ seems to be special
since the ``seed'' or ``source'' of our dispersion instability at
least partially lies in the soft plasmon emission process, but we see
no reason to associate $k_{th}$ with a characteristic charge density
wave instability. Similarly, the upper ($k_h^*$) and the lower
($k_l^*$) critical wavevectors where the mass divergence first
manifest itself (Figs.~\ref{fig:mk2D1} and \ref{fig:mk3D1}) could be
characteristic charge density wave vectors but we have seen no
theoretical indication of such a CDW instability, at least within our
RPA theory.

Anther possibility that has been much discussed in the literature
~\cite{khodel1, khodel2, nozieres, khveshchenko} is a superconducting
or fermionic condensation instability associated with the
quasiparticle energy degeneracy in the flat dispersion around $k_F$.
Such a superconducting instability is entirely different from the
usual Kohn-Luttinger superconducting instability (in some high angular
momentum channel) that is known to exist (with an exponentially low
superconducting transition temperature) in interacting electron
liquids. In our problem of dynamically screened Coulomb interaction,
such a superconducting instability could possibly arise from the
exchange of virtual plasmons (i.e.  plasmon-mediated
superconductivity) since emission of real plasmons is an underlying
mechanism for our dispersion instability. We have, in fact, little to
add to the existing discussion in the literature~\cite{khodel1,
  khodel2, nozieres, khveshchenko} on the issue of a superconducting
instability in the context of the dispersion instability except to
note that the transition temperature is likely to be rather low for
such a superconducting state.

We note in this context that Nozieres carried out~\cite{nozieres} a
penetrating and trenchant analysis of the original band-flattening
phenomenon introduced in Ref.~\onlinecite{khodel1}. While discussing
the instability, Nozieres~\cite{nozieres} was quite pessimistic about
the theoretical possibility of such an instability existing in a
realistic model that allows for screening. In fact, Nozieres concluded
that ``screening of a strong long range interaction is such that the
instability threshold cannot be reached''. Our work shows that this
conclusion of Ref.~\onlinecite{nozieres} is, in fact, too pessimistic
since our one-loop calculation is precisely an expansion in the
dynamically screened long-range Coulomb interaction.  Thus, the
effective mass divergence can certainly occur even when screening of a
strong long range interaction is explicitly incorporated in the theory
in contrast to Nozieres' conclusion in Ref.~\onlinecite{nozieres}.

A real theoretical concern is the possibility that the effective mass
divergence (and the dispersion instability) is just an unfortunate
artifact of our specific self-energy approximation, namely the
single-loop (leading-order in dynamically screened interaction) RPA
self-energy calculation. Although such a possibility can never be
ruled out theoretically (short of an exact treatment of the strongly
interacting electron liquid problem) we have argued~\cite{diverge}
elsewhere that this is unlikely to be the case here, i.e., there is
very good reason to believe that the dispersion instability at a low
electron density is a generic property of the strongly interacting
electron liquid (but the actual value of $r_s$ depends on the
approximation scheme). Without repeating the arguments already
discussed in Ref.~\onlinecite{diverge}, we point out that RPA (which
is a self-consistent field approximation, {\em not} a perturbative
expansion in $r_s$ although RPA does become exact in the high density
$r_s \to 0$ limit) works well (as compared with experimental data) at
metallic densities ($r_s \sim 3 - 6$) in 3D systems~\cite{book} and at
very low densities ($r_s \sim 10 - 20$) in 2D electron
systems~\cite{hwang}. Second, the RPA prediction for a ferromagnetic
instability in electron liquids turns out to be generically
``correct'', i.e. the most accurate QMC calculations also predict
ferromagnetic instabilities in 2D and 3D electron systems except that
RPA underestimates the critical $r_s$ for the electron liquid
ferromagnetic instability. This suggests that the RPA prediction for
the dispersion instability is also likely to be generically correct.
Third, the original dispersion instability, envisioned in
Ref.~\onlinecite{khodel1}, used a toy model (involving a rather
unrealistic long-range interaction) which is completely and
fundamentally different from our dynamically screened Coulomb
interaction model -- the fact that two completely different
interaction models arrive at very similar qualitative conclusions
about the dispersion instability (both in 2D and 3D) is again strongly
suggestive of the possible generic nature of the instability. Fourth
(and perhaps the most important of all), there are no general
theoretical arguments against such a dispersion instability, and
therefore, if the interaction is strong enough (i.e. $r_s$ large
enough), there is no theoretical reason for the instability {\em not}
to occur. This point becomes even stronger in the context of the
existence of such an effective mass divergence (``almost-localized
Fermi liquid'') in normal He-3 interacting via the short-range
interaction. We also note that, as shown in Ref.~\onlinecite{diverge},
the quasiparticle effective mass divergence occurs in many-body
approximations going beyond the RPA (e.g. the Hubbard
approximation~\cite{hubbard}), and the same is obviously true for the
dispersion instabilities. Thus the instability exists in self-energy
calculations beyond the one-loop approximation.

Finally, we discuss the experimental implications of our results.
First we note that the critical density ($r_s^* = 18$ (2D); $50$ (3D))
involved in the dispersion instability is extremely low for meaningful
experiments to be carried out in real systems. Also, our $T = 0$
theory does not say much about the finite temperature situation where
these experiments are necessarily carried out.  In fact, the
temperature scale at such low densities is likely to be small, and
therefore one may have to go to unrealistically low temperatures to
see any experimental consequences of the dispersion instability even
if the instability is real. It is important to emphasize that our RPA
values for critical $r_s^*$ are probably lower bounds on the true
$r_s^*$ which is likely to be larger than $r_s^* = 18$ (2D) and $50$
(3D).

With these caveats in mind, it is interesting to note that there have
been several recent experimental claims of the
observation~\cite{shashkin} of quasiparticle effective mass divergence
in Si MOSFET-based low-density interacting 2D electron system at $r_s
\sim 10$. These claims are, however, quite controversial, and in more
dilute 2D hole systems, where $r_s$ could reach as low as $20 - 30$,
no such mass divergence has been reported. We believe that these
recently reported effective mass divergence claims in Si MOSFETs are
most likely not connected with the dispersion instability discussed in
our work. This is particularly true in view of the fact that the
reported effective mass divergence does {\em not} seem to be a
dispersion instability, and furthermore, the experimental critical
$r_s^*$ ($\sim 10$) is far too low compared with the theoretical
finding ($\sim 18$).  In fact, we expect our theoretical $r_s^*$ to be
much lower than the RPA value ($r_s^* \sim 18$ in 2D) predicted in our
work once quasi-2D finite width effects and many-body effects going
beyond RPA are included in the theory. In addition, the semiclassical
experimental technique of using Dingle fits to temperature-induced SdH
amplitude decay~\cite{shashkin} in extracting the quasiparticle
effective mass is highly suspect in such a strongly interacting
quantum system. The experiments claiming the effective mass divergence
also completely ignored the strong temperature
dependence~\cite{finiteT} of the quasiparticle effective mass which
certainly invalidates the semiclassical Dingle fitting procedure
employed in the experiments in obtaining the quasiparticle effective
mass. We therefore conclude that the experimental consequences of the
dispersion instability in low-density interacting electron liquids
remain an interesting open challenge for the future.

We acknowledge fruitful discussions with M. V. Zverev and V. A.
Khodel. This work is supported by the US-ONR and NSF.


\bibliography{dispinst}

\end{document}